# Photometric and spectroscopic analysis of Comet 29P/Schwassmann-Wachmann 1 activity

Oleksandra V. Ivanova [a,*], Igor V. Luk'yanyk [b], Nikolay N. Kiselev [a], Viktor L. Afanasiev [c], Enos Picazzio [d], Oscar Cavichia [e], Amaury A. de Almeida [d], Sergei M. Andrievsky [f]

[a] Main Astronomical Observatory, National Academy of Sciences of Ukraine, Goloseevo, Kyiv 03680 Ukraine
[b] Astronomical Observatory, Taras Shevchenko National University of Kyiv, Observatorna str. 3, Kyiv 04053, Ukraine
[c] Special Astrophysical Observatory of the Russian AS, Nizhnij Arkhyz 369167, Russia
[d] Universidade de São Paulo, Instituto de Astronomia, Geofísica e Ciências Atmosféricas, Departamento de Astronomia, Rua do Matão, 1226; Cidade Uni-versitária, CEP 05508-900, São Paulo, SP, Brasil
[e] Instituto de Física e Química, Universidade Federal de Itajubá, Av. BPS, 1303; Pinheirinho, CEP 37500-903, Itajubá, Brazil
[f] Department of Astronomy and Astronomical Observatory, Odessa National University, Shevchenko Park, 65014 Odessa, Ukraine

We carried out photometric and spectroscopic observations of comet 29P/Schwassmann-Wachmann 1 at the 6-m BTA telescope (SAO RAS, Russia) and the 1.6-m telescope of the National Laboratory for Astro-physics (LNA, Brazil) on February 20, 2012, and on May 31, 2011, respectively. The spectra revealed the presence of $CO^+$ and $N_2^+$ emissions in the cometary coma at a distance of 5.25 AU from the Sun. The ratio $[N_2^+]/[CO^+]$ within the projected slit is 0.013. The images obtained through BVR filters showed a bright, dust coma, indicating a high level of activity. We estimated a colour index and a colour excess for the comet. The parameter $Af\rho$, which is used as an indicator of a cometary activity, was measured to be $2584 \pm 50$ cm in the reference optical aperture of $\rho = 10^4$ km. The dust production constituted 33 kg/s and $9.3 \cdot 10^3$ kg/s, it was obtained using different methods. We also investigated the morphology of the comet using image enhancement techniques and found two jets in the coma.

## Introduction

Comet 29P/Schwassmann-Wachmann 1 (hereafter SW1) was discovered in 1927 on photographic plates by German astronomers A. Schwassmann and A. Wachmann. The semi-major axis of its orbit, eccentricity, and orbital period are 6.01 AU, 0.042, and 14.66 yrs, respectively. The nucleus radius estimate exceeds that of a typical short-period comet. Visible photometry (Meech et al., 1993) and analysis of the thermal emission (Cruikshank and Brown, 1983; Stansberry et al., 2004) yields an estimate of the radius of SW1 as 15–44 km.

SW1 is regarded to be a comet of the Centaurs class. Such comets could be runaways from the Kuiper belt. SW1 manifests an episodic flare activity during the long period of observations. However, Jewitt (1990) pointed out that the SW1 coma never disappears completely. Typically, 2–3 outbursts per year occur, but the frequency of the events is poorly known. Generally, the intensity of the outburst is proportional to the product of the gas production and lifetime of molecules. In other words, the light flare can be caused either by the gas production increase (Trigo-Rodríguez et al., 2008), or by the change of the lifetime of molecules caused by the solar UV radiation (Kravtsov and Luk'yanyk, 2010). Since comet SW1 is located at the distance 6 AU, i.e., outside the zone of the water sublimation, it is obvious that mechanisms that are based on the change of the molecules' lifetime should be discarded.

As a rule, the outburst activity caused by the gas production is believed to be connected with CO. Neutral CO that was detected in the coma of comet SW1 is thought to be a primary driver gas for the observed dust and gas morphological features (Senay and Jewitt, 1994). But the release of trapped gases owing to the exo-thermic phase transformation of water ice from amorphous to crystalline state is found to be insufficient to generate the observed activity. Therefore, the question of the SW1 activity is still open.

Optical spectroscopy reveals the presence of the $CO^+$ ions and CN radicals in the coma of comet SW1 (Cochran et al., 1991a,b 1980, 1982; Cook et al., 2005; Larson, 1980; Korsun et al., 2008). The emissions were observed in both a quiescent state and outburst. Finally, the origin and orbital evolution of SW1 is not yet fully understood.

## Observation and reductio

### Photometry

The observations of comet SW1 were made at the 6-m tele-scope of SAO RAS on February 20, 2012. At that time, the helio-centric and geocentric distances of the comet were 6.26 and 5.49 AU, respectively. The phase angle of the comet was 6.2°. We  observed the comet in period of moderate outbursts (Trigo-Rodríguez et al., 2012; Paganini et al., 2013). The observations were made using the universal SCORPIO-2 focal reducer, mounted in the primary focus of the 6-m telescope (Afanasiev and Moiseev, 2011). The SCORPIO-2 device with an E2V 42-90 CCD detector was used in photometric mode. The dimension of the image was $1024 \times 1024$ pixels (binning $2 \times 2$) and the scale was $0.18''/\text{pix}$. The full field of view of the CCD is $6.1' \times 6.1'$.

The photometric data of comet SW1 were obtained in Johnson–Cousins broadband system BVR. Seeing was quite stable during the observations, with a mean value of $1.0''$ (FWHM) in the R filter. To increase the signal-to-noise ratio for the image analysis, we decided to co-add the images obtained for each filter.

To reduce the redundancy and improve the signal-to-noise ratio, the images were saved with binning equal to 2, hence the operating image scale is $0.36''/\text{pix}$. The twilight morning sky emission was registered for the flat fields. Standard bias subtrac-tion and flat field reduction for all the data were performed. We used routines of the IDL library (Goddard Space Flight Centre) to calculate the sky background count (Landsman, 1993). For more details about observational data, see Table 1. We obtained the integrated magnitude of the comet as a function of aperture. The spectrophotometric standard star BD +25°4655 was observed in order to perform absolute calibration (Oke, 1990).

### Spectral mode

The spectrophotometry observations were made at the 1.60 m telescope of the National Laboratory for Astrophysics (LNA, Brazil) during May 31, 2011, following the same instrumental setup and data reduction described in Cavichia et al. (2010). Five spectra were obtained during the observation night. Each of the cometary spectra was taken with one exposure of 2400 s and a long slit of 1.5 arcsec width. A Cassegrain Boller & Chivens spectrograph was used with a 300 l/mm grid, which provides a reciprocal dispersion of 0.24 nm/pixel. A Marconi CCD 42-40 was used with an operat-ing image scale of $0.56''/\text{pix}$ and a pixel size of $13.5 \times 13.5 \, \mu\text{m}$. The non-sideral tracking was performed by applying manually offsets in order to keep the position of the moving target in the slit during the exposure. The spectrophotometric standard stars CD-32°9927 and LTT 6248 from Hamuy et al. (1992, 1994) were observed to improve the flux calibration. These stars were observed with a long slit of 7.5 arcsec width, allowing a more precise flux calibra-tion. Helium–argon arcs were taken immediately after each sci-ence spectra in order to perform wavelength calibration.

Data reduction was performed using the IRAF package, fol-lowing the standard procedure for long slit spectra: correction of bias, flat-field, extraction, wavelength and flux calibration. Sky subtraction and the extraction of the cometary spectrum were performed using the IRAF task "apall". For the spectra, the skylines were subtracted using the line fluxes at a "safe" distance on both sides of the comet spectrum, where no coma light was present. A Chebyshev function of order 3 was used in order to fit the sky background. The cometary spectrum was extracted by defining an aperture width of 351 pixels around the centre of the distribution, which includes the coma light. In this way, it was possible to extract the cometary spectrum properly corrected by the sky level subtraction.

The comet showed high activity during our observation. Trigo-Rodríguez et al. (2011) presented results of photometric monitor-ing, which showed the most noticeable activity of the comet during May 2011, with the last outburst on May 22, 2011.

## Observational results

### Photometric data analysis

The BVR magnitudes of the comet were obtained during the high quality photometric night by performing the absolute flux calibration using the spectrophotometric standard star BD +25°4655. We also calculated reduced R magnitudes using the following relation:

$$m_R(1,1,0) = m_R - 5 \cdot \log_{10}(r \cdot \Delta) - \Phi(\alpha), \tag{1}$$

where $r$ and $\Delta$ are the heliocentric and geocentric distances in AU, respectively, $\Phi(\alpha)$ is the so-called phase function correction at a phase angle $\alpha$. The shape of this function depends upon a variety of properties of dust particles, including the size distribution, composition, and albedo. We used tabulated values of $\Phi(\alpha)$ nor-malized at a phase angle of 0° (http://asteroid.lowell.edu/comet/dustphaseHM_table.txt); the method is described in Schleicher et al. (1998) and Schleicher (2007).

The colour excesses were obtained using the next relationship:

$$CE = C - [m_{sun}(\lambda_1) - m_{sun}(\lambda_2)], \tag{2}$$

where $C = m_c(\lambda_1) - m_c(\lambda_1)$ is the observed colour index of the comet, and $[m_{sun}(\lambda_1) - m_{sun}(\lambda_2)]$ is the colour index of the Sun. The solar colour indexes $B–V$ and $V–R$ are 0.65 and 0.35, respectively (Holemberg et al., 2006). The distant comet SW1 is an emission-rich comet. The spectra of the comet (in this work) show the presence of the emissions of CN, $CO^+$, and $N_2^+$. Broadband BV filters cover those emission bands of cometary molecules. We estimated the contribution of those emissions to the total flux through filter B from spectral observations (Section 3.3). This contribution is less 2%. The main contribution is the continuum. This value is the upper limit for other filters, because the B filter covers most of the emission features of the SW1 spectra. In addition, our estimate shows that colour indexes of the comet are redder than the solar ones, assuming that the emissions do not influence the dust colour the comet. The integral magnitude as a function of the aperture size, reduced R magnitude, and colour excesses are shown in Table 2 and compared with other Centaurs in Fig. 1.

**Table 1**
Log of the observations for comet 29P/Schwassmann-Wachmann 1.

| Date observ. (UT) | Exptime (s) | $r$ (AU) | $\Delta$ (AU) | $\alpha^a$ (°) | $z^b$ (°) | Type | Info$^c$ |
|---|---|---|---|---|---|---|---|
| 2011 | | | | | | Spec. | |
| May. 31.89 | 2400 | 6.25 | 6.17 | 9.3 | 23 | | R=900 |
| 2012 | 600 | | | | | Phot. | B |
| Feb. 20.59 | 450 | 6.26 | 5.49 | 6.2 | 58.2–59.4 | | V |
| | 120 | | | | | | R |

$^a$ Phase angle.
$^b$ Zenith distance.
$^c$ Spectral resolution and using filters.

| $\rho$ (arcsec) | $m_R$ (mag) | $C_{B-V}$ (mag) | $C_{V-R}$ (mag) | CE (mag) | | $m_R(1,1,0)$ (mag) | $Af\rho_R$ (cm) |
|---|---|---|---|---|---|---|---|
| | | | | $B-V$ | $V-R$ | | |
| 1.1 | $16.65 \pm 0.01$ | $0.79 \pm 0.04$ | $0.64 \pm 0.03$ | 0.14 | 0.12 | $12.55 \pm 0.01$ | $2406 \pm 121$ |
| 2.2 | $15.82 \pm 0.01$ | $0.75 \pm 0.03$ | $0.62 \pm 0.02$ | 0.10 | 0.10 | $10.22 \pm 0.01$ | $2584 \pm 50$ |
| 3.3 | $15.39 \pm 0.01$ | $0.73 \pm 0.03$ | $0.62 \pm 0.02$ | 0.08 | 0.10 | $8.87 \pm 0.01$ | $2656 \pm 30$ |
| 6.6 | $14.48 \pm 0.01$ | $0.72 \pm 0.03$ | $0.62 \pm 0.02$ | 0.07 | 0.10 | $6.43 \pm 0.01$ | $2779 \pm 18$ |

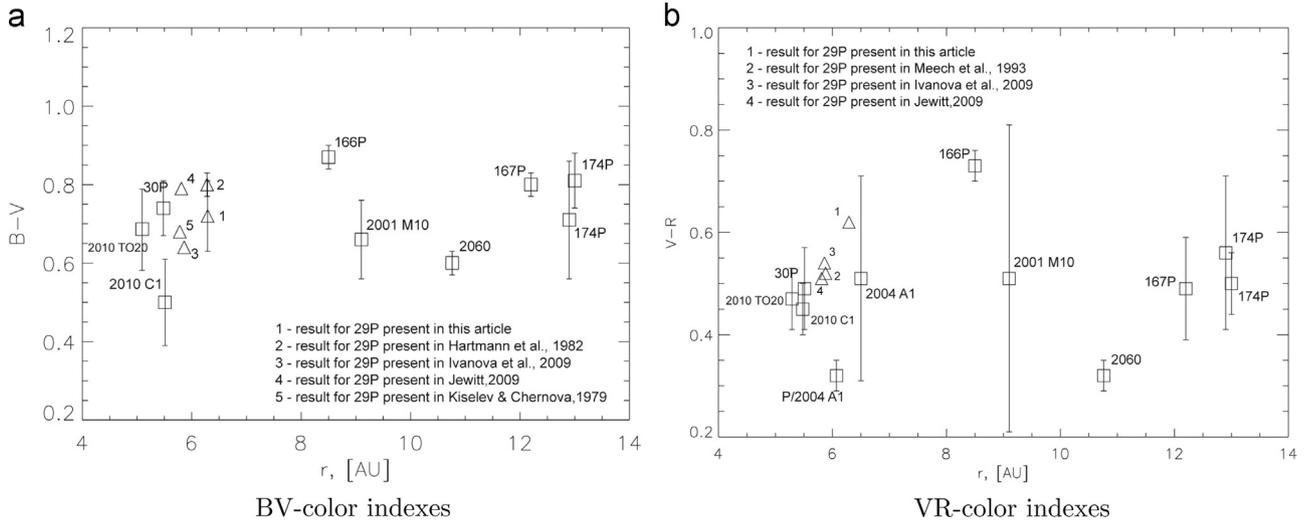

BV-color indexes     VR-color indexes

**Fig. 1.** Colour indexes for the comet 29P/Schwassmann-Wachmann 1 (open triangles) compared to values derived for active Centaurs (open square). The values of the colour indexes obtained from literature: Epifani et al. (2014), Lacerda (2012), Jewitt (2009), Bauer et al. (2008), Rousselot (2008), Romon-Martin et al. (2003): (a) BV-colour indexes and (b) VR-colour indexes.

The last analyse of SW1 spectra (Cochran et al., 1980; Korsun et al., 2008) and our observations (see Section 3.3) of the cometary spectra in the R domain are almost free from emission lines. It allowed us to use the R magnitudes for calculation of the aperture-independent parameter Af$\rho$ (A'Hearn et al., 1984). The results are presented in the last column of Table 2.

For the estimation of the dust mass production rate, we used two methods. First, we used the method presented in the paper of Jewitt (2009). The apparent magnitude $m_d$ of the coma was estimated in the annulus between the aperture radii $\rho_1 = 2.2$ arcsec and $\rho_2 = 3.3$ arcsec. The relationship between apparent magnitude $m_d$ and $p(\lambda)$ (the geometric albedo of the dust grains), and $C_d$ (the effective cross-section of the dust particles, captured by the annular aperture, projected on the celestial sphere), is expressed by the ratio given by Russel (1916) and Jewitt (2009).

$$p(\lambda) \cdot C_d = \frac{2.25 \cdot 10^{22} \cdot \pi \cdot r^2 \cdot \Delta^2 \cdot 10^{-0.4(m_d - m_{sun})}}{\Phi(\alpha)}, \qquad (3)$$

where $m_{sun}$ is the magnitude of the Sun ($-26.098$, $-26.74$, and $-27.094$ for the B, V, R bands, respectively). For our calculations, the geometric albedo was set to be equal to 0.1, which is consistent with measurements of active comets in backscattering geometry (Kolokolova et al., 2004).

The obtained dust cross-section value is related to the dust mass $M_d$ by means of the dust size distribution. The mass of the dust can be represented by the following equation:

$$M_d = \tfrac{4}{3} \cdot \rho_{dust} \cdot (a_{min} \cdot a_{max})^{0.5} \cdot C_d, \qquad (4)$$

where $\rho_{dust}$ is the grain density that does not depend on radius of the dust grains, and is equal to 1000 kg/m³, $a_{min}$ and $a_{max}$ are the minimum and maximum radii of the grain particles, respectively, which we adopt to be $10^{-1}$ and $10^3$ µm (Grün et al., 2001). Following Jewitt (2009), the grain residing time in the annulus defined by $\rho_1$ and $\rho_2$ is:

$$\tau(\rho) = 1.5 \times \frac{10^{11} \cdot \Delta \cdot (\rho_2 - \rho_1)}{v_0} \cdot \left[\frac{r}{r_0}\right]^{1/4}, \qquad (5)$$

where $v_0 = 550$ m/s is the velocity at reference distance $r_0 = 5$ AU (Biver et al., 2002), $\rho_1$ and $\rho_2$ are the aperture radii in radians. To compare our results with those obtained by Jewitt (1990, 2009) for comet SW1, we used the same albedo and apertures.

The second method we used for estimating the dust mass production rate is based on Newburn and Spinrad (1985) and Weiler et al. (2003). We refer to Korsun et al. (2014) and Rousselot et al. (2014) for further details on this method. We adopted the particle size distribution function in the form of $f(a) \sim a^{-3.3}$ (Fulle, 1992; Moreno, 2009). The expression for the dust density was taken from Newburn and Spinrad (1985). The range of dust outflow velocity (10–20 m/s) was based on results obtained from numerical modelling of the dust environment of this comet and presented in the articles of Fulle (1992) and Moreno (2009). The estimated dust-loss rates of the comets and those taken from literature are listed in Table 3.

*Morphology of the cometary coma*

The comet SW1 typically demonstrates long-term outburst activity in the form of dust jets. Our observations of the comet showed a high activity within the coma. To select the weak contrast structures (jets) within the images of the dust coma, we used the special software Astroart (http://www.msb_astroart.com/), which is provided with a number of digital filters. The detailed

**Table 3**
Model results of dust mass-loss rate for the comet 29P/Schwassmann-Wachmann 1.

| $r$ (AU) | $\Delta$ (AU) | $\alpha$ (deg) | $m_d$[a] | $M_d$ (kg) | | $v$ (m/s) | | $dM_d/dt$ (kg/s) | | $A_r$ | Data |
|---|---|---|---|---|---|---|---|---|---|---|---|
| | | | | A[b] | B[c] | A | B | A | B | | |
| 6.29 | 5.49 | 6.2 | 16.6 | $7.87 \cdot 10^7$ | $4.0 \cdot 10^4$ | 520 | 11–23 | $9.3 \cdot 10^3$ | 33 | 0.1 | Our observation |
| 5.79–5.91 | 4.81–5.89 | 1.3–10 | 17 | $6 \cdot 10^5$ | | – | | 10 | | 0.04 | Jewitt (1990) |
| 5.81 | 4.99 | 6.2 | 17.0 | $3.84 \cdot 10^7$ | | 529 | | $5.1 \cdot 10^3$ | | 0.1 | Jewitt (2009) |
| 5.72 | 5.29 | 9.6 | 17.5 | – | | 18.8 | | 500–2000 | | 0.1 | Moreno (2009) |
| 5.77 | 5.12 | 8.2 | – | – | | 600 | | 1–20 | | 0.1 | Fulle (1992) |

[a] Coma magnitude (in the R filter) in the projected annulus between $\rho = 2.2$ arcsec and $\rho = 3.3$ arcsec.
[b] Calculation were made using the equation from the paper by Jewitt (2009).
[c] Estimated dust production rate of the comet nucleus taking into account the discussion given.

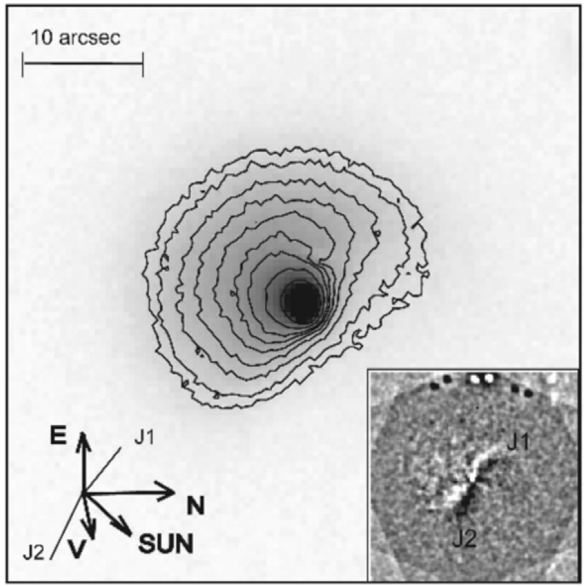

**Fig. 2.** The isophotes and two jets in the dust coma of the comet 29P/Schwassmann-Wachmann 1 revealed after processing images of the comet by the Larson & Sekanina filter obtained on February 20, 2012. The directions to the North (N), to the East (E), to the Sun and moving direction (V) of the comet are indicated.

information on the digital filters can be found in Ivanova et al. (2012).

To eliminate false details from the images, each of the filters was separately applied to each image. Such a technique of enhancing the structures in the coma was successfully used by Manzini et al. (2007) for comet C/2002 C1 Ikeya-Zhang, Korsun et al. (2008, 2010) for distant comets, and by Ivanova et al. (2012) for comet SW1.

To our images, we first apply the Larson & Sekanina filter (Larson and Sekanina, 1984). After this filtering, we selected jets structures. To improve the presentation of the structure (more contrast), we apply two additional filters: the unsharping mask and the Gaussian blur. The final result is presented in Fig. 2, showing two distinct jets in comet SW1 in February 2012. The jets can be isolated on the background of the bright structureless coma. The jets are directed outward from the photometric centre of the comet in perpendicular directions relatively to the Sun. Nevertheless, it should be noted that digital filters are non-linear. Therefore, there exists some possibility of false structures or some shift of the spatial position of the real jets (Fig. 3).

*The spectroscopic data analysis*

The resulting spectrum of SW1 is shown in Fig. 2. The short wavelength region is dominated by the rather strong noise. For the sake of comparison, we reproduce in this figure the solar spectrum in relative intensities (Neckel and Labs, 1984). The solar spectrum was convolved with the corresponding Gaussian profile to reach spectral resolution of the cometary spectrum. Qualitative comparison of the cometary and solar spectra indicates the presence of weak molecular emissions in the comet and some differences of the flux distribution between both spectra.

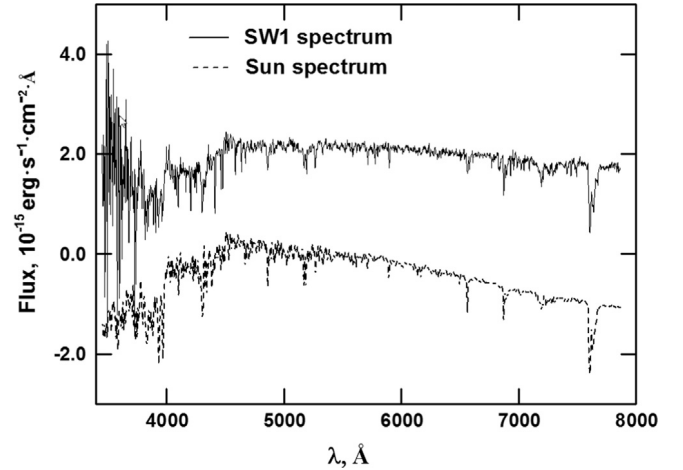

**Fig. 3.** Energy distribution in the spectrum of the comet and arbitrarily normalized solar spectrum.

The spectral dependence between continua of the cometary and solar spectra, which is caused by the different efficiencies of the solar radiation scattering by the cometary dust at different wavelengths, can be estimated after dividing the cometary spectrum by the solar one.

Using the median filter with wide window, we obtained a smoothed curve that reproduces spectral dependence of the solar radiation scatter efficiency caused by the cometary dust particles. Our result is given in Fig. 4. From this figure, one can see that there is a nonlinear increase with the wavelength of the scattering efficiency. In the short wavelength region (less than 3800 Å), the scattering efficiency was not determined because of the high noise present in the cometary spectrum.

The spectral reflectance is $S(\lambda) = F_c(\lambda)/F_s(\lambda)$. Here, $F_c(\lambda)$ is the cometary continuum, and $F_s(\lambda)$ is a scaled spectrum of the Sun. The normalized reflection ability $S'$ can be described as (Jewitt and Meech, 1986):

$$S'(\lambda_1, \lambda_2) = \frac{\frac{dS}{d\lambda}}{S_{mean}}, \qquad (6)$$

where $dS/d\lambda$ is the rate of change of the reflectivity with respect to wavelength in the region from $\lambda_1$ to $\lambda_2$ and $S_{mean}$ is the mean

reflectivity in the observed wavelength range:

$$S_{mean} = N^{-1} \sum S_i(\lambda). \tag{7}$$

The $S'(\lambda_1, \lambda_2)$ is expressed in percent per $10^3$ Å, and provides a convenient measure of the grain colour. For comparison with the earlier results, let us present the linear approximation of our results performed by the least square method for the most typical spectral regions that are defined by the effective wavelengths of the cometary filters used for the continuum registration [BC (4430 Å), GC (5260 Å) and RC (6840 Å)], see Schleicher et al. (2004): $11.4 \pm 2.3\%$ per $10^3$ Å for the range 4430–5260 Å, $17.9 \pm 5.6\%$ per $10^3$ Å for the range 5260-6840 Å and $14.8 \pm 4.8\%$ per $10^3$ Å for the range 4430–6840 Å. The values obtained for the normalized spectral gradient for ranges 4430–5260 Å and 5260-6840 Å are comparable within the error bars to the redder wavelength region. This result does not allow an unambiguous conclusion about the normalized spectral gradient behaviour with increasing wavelength.

By multiplying the solar spectrum by the scattering efficiency we can get an estimate of the real cometary continuum. After subtraction of the obtained continuum from the observed spectrum, the fluxes of molecular emissions are determined.

To increase the signal/noise ratio all the spectra were co-added. Part of the spectrum with a wavelength of less than 3800 Å was not considered as it was very noisy.

Spectroscopic observations of distant objects in the Solar System with cometary activity are not numerous. Only a few emissions of gaseous species have been observed in the optical domain beyond 5 AU. Molecular emissions have only been detected in the peculiar periodic comet 29/Schwassmann-Wachmann 1 (CO$^+$, CN), giant comet Hale-Bopp (CN, C$_3$), and Centaur Chiron (CN).

The linear spectrum of the comet is shown in Fig. 5. Despite the rather high noise level in the blue region, we nevertheless succeeded to confidently register some spectral details that may be associated with molecular emissions from the coma.

An identification of these details was made by means of comparison with laboratory and calculated molecular spectra in the same spectral region.

Since the resolution of our spectra is similar to that of Korsun et al. (2008), we used the calculated template spectra of CN, N$_2^+$ and CO$^+$ molecules from that work (Table 2). The extracted molecular spectrum is displayed in Fig. 5.

The strongest features which extend along the whole observed spectral window are the omet-tail bands of CO$^+$. The lines (2,0), (3,0), (2,0), (1,0), (5,1), (3,1), (2,1), (4,2), (3,2), (0,0), and (1,1) of the vibrational transitions of CO$^+$ (A$^2\Pi$–X$^2\Sigma$) band system are clearly seen in Fig. 5. Two weak bands, (0,1) and (1,2), which belong to the (B$^2\Sigma$–A$^2\Pi$) system (Baldet & Johnson) of the CO$^+$ ion, are detected as well. The N$_2^+$(0, 0) of the (B$^2\Sigma$–X$^2\Sigma$) electronic system is shown too.

The [N$_2^+$]/[CO$^+$] ratio is important in our understanding of the Solar System nebula formation. To estimate that ratio we used integrated intensities of the CO$^+$(2, 0) and N$_2^+$(0,0) bands. The column density is defined by

$$N = \frac{L}{g_{v'v''}}. \tag{8}$$

where $N$ is the column density, $L$ is the integrated band intensity and $g_{v'v''}$ is the excitation factor. We used excitation factors of $7.0 \times 10^{-2}$ photons s$^{-1}$ mol$^{-1}$ for the N$_2^+$(0,0) band (Lutz and Womack, 1993) and $3.55 \times 10^{-3}$ photons s$^{-1}$ mol$^{-1}$ for the CO$^+$(2,0) band (Lutz and Womack, 1993; Magnani and A'Hearn, 1986). Finally, the ratio can be derived using the following expression (Cochran et al., 2000):

$$\frac{N_2^+}{CO^+} = \frac{g_{CO^+}}{g_{N_2^+}} \cdot \frac{L_{N_2^+}}{L_{CO^+}} \tag{9}$$

If only the (2,0) band column density of CO$^+$ is used, then [N$_2^+$]/[CO$^+$] should be equal to 0.013. But this is only the upper limit for this ratio. Fig. 6 shows the calculated spectral profiles CO$^+$(5,1) and N$_2^+$(0,0). Those profiles were calculated in view of the spectral resolution of the observed spectrum of comet SW1. The N$_2^+$ calculated spectrum was made using the LIFBASE software (Luque and Crosley, 1999) and the CO$^+$ one using data from Kim (1994). CO$^+$(5,1) bands show double peaks. The second peak coincides with N$_2^+$. We estimated the available contamination of N$_2^+$ by CO$^+$(5,1). This contamination achieves 19%. Therefore, [N$_2^+$]/[CO$^+$] should be equal to 0.01. This result agrees well with Arpigny's estimate of the intensity ratio for [N$_2^+$]/ [CO$^+$] equal to 0.011 for comet SW1 (Cochran et al., 2000; Korsun et al., 2008).

## Discussions

The comet-centaur 29P/Schwassmann-Wachmann 1 is an interesting object with near circular orbit and outburst activity.

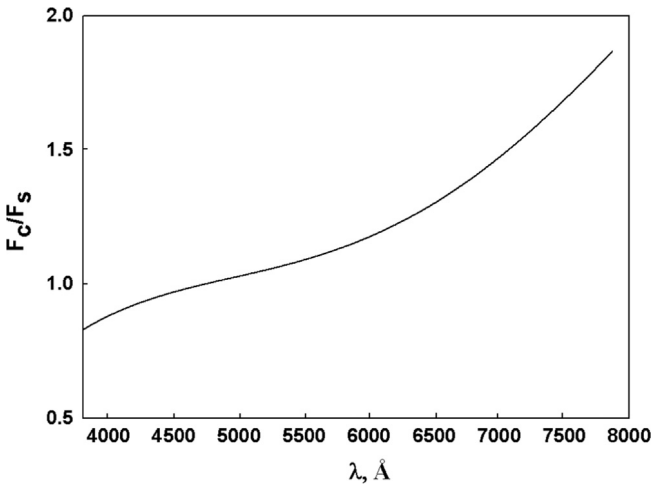

**Fig. 4.** Reddening $S(\lambda)$ of the observed spectrum relative to the solar spectrum.

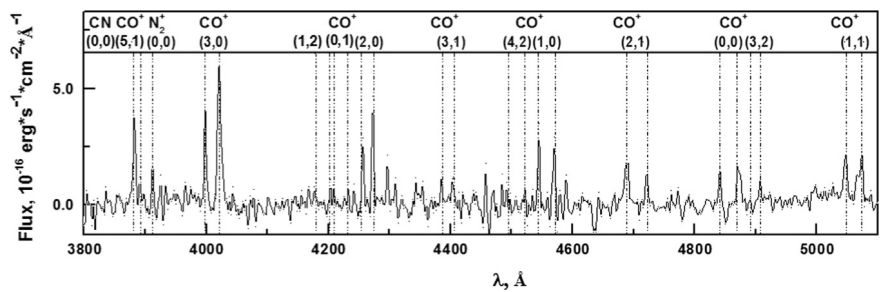

**Fig. 5.** Molecular emissions identified in the observed spectrum of comet SW1.

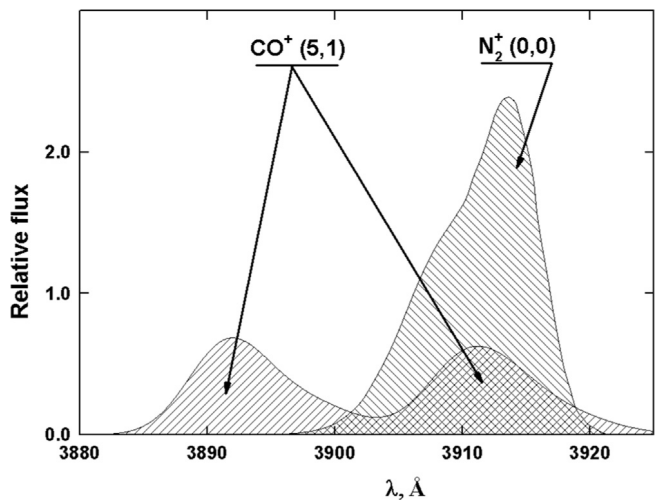

**Fig. 6.** Calculated spectroscopic profiles of the $N_2^+$ and $CO^+(5,1)$ bands.

Previous observations of the comet with different methods showed $CO/H_2O > 1.6$ (Biver et al., 2007; Gunnarsson et al., 2008), $CO/H_2O \sim 10$ (Bockelée-Morvan et al., 2010), and $CO/H_2O > 4.7$ (Ootsubo et al., 2012), showing that the comet is CO rich. Last ground-based infrared observations of the comet (Paganini et al., 2013) confirm that CO can be the main driver, which controls the activity of comet SW1. Hill et al. (2001) argue that comets, formed early in nebular history, should be rich in CO, $CO_2$, $N_2$, and amorphous ice. The spectra of CO-rich comets show the significant predominance of $CO^+$, strong emissions of CN and $N_2^+$, and measurable $C_3$ features. Actually the spectra of SW1 are characterized by the predominance of $CO^+$ and $N_2^+$ emissions (Korsun et al. (2008); this work). Meanwhile, Cochran et al. (2000) and Cochran (2002) confirm the identification of $CO^+$ only in the spectra of the comet. Possible mechanisms of ionization of the parent CO in cometary coma at large heliocentric distance are not fully understood, too. The results of a new scenario consideration that was invoked to explain an apparent $N_2$ deficiency in comets proposed Mousis et al. (2012) can be used for comets where no detection of $N_2^+$ is made. Recent results obtained from the ROSINA mass spectrometer onboard of ROSETTA also detected molecular nitrogen in the nucleus of comet 67P/Churyumov-Gerasimenko. All of these results make it possible to comment on the formation scenario of the Solar System. The sublimation temperatures of CO and $N_2$ are 25 K and 22 K, respectively. Recent laboratory experiments indicate that the ice grains, which accumulated to produce comet nuclei, were formed by freezing water vapour at about 25 K (Bar-Nun et al., 2007; Notesco and Bar-Nun, 2005; Notesco et al., 2003). From one side the CO rich comet SW1 must be formed in the outer regions of the disk after collapse of the nebular cloud (in cold regions) based on the analysis proposed by Paganini et al. (2012), but new Spitzer observation show the presence of crystalline silicates in the coma (Stansberry et al., 2004; Kelley et al., 2009), which must have experienced strong thermal processing (hot regions) very near the young Sun. Perhaps such features may be related to the place and conditions of formation of comets in Solar System.

Photometric studies of comet objects show differences of the chemical composition of comets and its connection with places of formation (A'Hearn et al., 1995). We detected CN emission in the spectrum of SW1, also. The distant detection ($\geq 4$ AU) of neutral gas species has been made in a very few comets (SW1, C/2002 VQ94 (LINEAR), C/1995 O1 (Hale-Bopp), 1P/Halley) and in Centaur (2060) Chiron. In the literature a few sources of CN in comets have been proposed. The outflow of the parent of CN in comets can be driven by sublimation of CO or $CO_2$ ices. It is known that HCN was present in 1P/Halley (Schloerb et al., 1987) and it is presumably one of several parents of CN in that comet. Another source of CN in 1P/Halley was the direct release from the distributed grains in the coma, probably grains of CN-bearing polymers (A'Hearn et al., 1986). Model of the activity of SW1 shows that icy grains can serve as an extended source of more volatile species (Gunnarsson, 2003) and a source of $H_2O$ from the analysis of Hershel observations of the comet (Bockelée-Morvan et al., 2010a). Given the results of observations of long lasting outburst activity of SW1, the scenario of an outburst to explain the detection of CN in the comet SW1 is more appropriate, but we cannot rule out the steady-state scenario.

Comets can display different dust properties. Dust colour gives indications of the wavelength dependence of the refractive indices. Observations of solar light scattered by solid particles in cometary comae depend on the observational conditions (e.g., phase angle, wavelength) and on the particles physical properties (e.g., complex refractive indices and composition, size, shape, geometric albedo). Variations in the size of the grains produce variations in the colour of the dust. Some authors indicate variations of the colour with the heliocentric distance, but the trend is not clear. The colour changes progressively from red to neutral as the wavelength increases from the optical to the near-infrared and for the optical range the mean value equals 13%, and changes from 5% to 18% (Jewitt and Meech, 1986). On average, the reddening is about 16% per $10^3$ Å for cometary dust (A'Hearn et al., 1995). For five Jupiter-family comets, the average value is 19% per $10^3$ Å (Hadamcik and Levasseur-Regourd, 2009). The colour of the Jupiter-family comets dust seems to be redder than the colour of other comets but the difference is in the error bars. We have calculated the average value of the normalized spectral gradient for Centaurs (in Table 5 from Peixinho et al., 2015). The Centaurs (as the Jupiter-family comets) are often faint objects and the dust colour measurements suffer large uncertainties. The value of the normalized spectral gradient for the Centaurs is $21.3 \pm 1.4$ per $10^3$ Å. We found that the mean value of the normalized spectral gradient for SW1 is $14.7 \pm 4.2\%$ per $10^3$ Å. The values of the normalized gradient of SW1 obtained for spectral range 4430–5260 Å and 5260–6840 Å do not allow conclusions about its behaviour with increasing wavelength (comparable results within the error bars along the redder wavelength region).

The observed change in the colour of the comet SW1 (Fig. 1) can be related to its degree of activity in the observed period. SW1 appears to have asymmetrical coma with jet features, beyond the distance at which water ice sublimates. Also the comet shows outburst activity with correlation to the number of jets and dust productivity (Ivanova et al., 2009). Observations presented by Trigo-Rodríguez et al. (2008) show that the SW1 coma is continuously supplied with fine dust. In periods when the comet is faint, it retains its diffuse appearance, but coma never completely disappears. Such a continuous dust outflow is also supported by Spitzer observations (Stansberry et al., 2004), since it indicates that a low-level activity is also maintained between outbursts.

## 5. Conclusions

1. The results obtained for the colour show that the cometary continuum is essentially more reddish than solar, and this can be caused by activity of the comet during its photometric observations (February 20, 2012).

2. The dust production on the date of observation (February 20, 2012) constituted 33 kg/s and $9.3 \cdot 10^3$ kg/s, it was obtained using different methods. Our results are comparable with estimate for comets that are at similar heliocentric distances of about 6 AU.

3. Using digital filters, we succeeded to isolate 2 "jet-like" features in the cometary coma, supporting the high activity in the period of observations on February 20, 2012.

4. Spectral dependence of the light scattering by the cometary dust is obtained from the spectral observations (May 31, 2011). The mean value of the normalized spectral gradient is $14.7 \pm 4.2\%$ per $10^3$ Å.

5. We detected numerous lines of $CO^+$ as well as the $N_2^+(0,0)$ line of the $(B^2\Sigma–X^2\Sigma)$ system, suggesting that the comet was formed in a low temperature (about 25 K) environment.

6. The value of $[N_2^+]/[CO^+]$ is equal to 0.01 for comet SW1.

## Acknowledgements

The observations were performed thanks to the support of the Schedule Committee for Large Telescopes (Russian Federation). The spectroscopic data were obtained from observations made at the Observatório Pico dos Dias/LNA - Brazil. We thank D.C. Boice for useful comments on the manuscript. And we are grateful to our reviewers for the thoughtful reviews, whose comments and suggestions greatly aided this paper.